\documentclass[twocolumn, superscriptaddress]{revtex4-1}
\usepackage{bm}
\usepackage[pdftex]{graphicx}
\usepackage{color}
\usepackage{soul}
\usepackage{xcolor}
\usepackage{amsmath}
\usepackage{amssymb}
\usepackage{url}
\usepackage[utf8]{inputenc}

\begin{document}

\title{A general statistical model for waiting times until collapse of a system}

\author{Vivianne Olgu\'in-Arias}
\email{vivianne.olguin.a@gmail.com}
\affiliation{Comisión Chilena de Energía Nuclear, Casilla 188-D, Santiago, Chile}

\author{Sergio Davis}
\affiliation{Comisión Chilena de Energía Nuclear, Casilla 188-D, Santiago, Chile}
\affiliation{Departamento de F\'isica, Facultad de Ciencias Exactas, Universidad Andres Bello. Sazi\'e 2212, piso 7, Santiago, 8370136, Chile.}

\author{Gonzalo Gutiérrez}
\affiliation{Grupo de Nanomateriales, Departamento de F\'{i}sica, Facultad de Ciencias, Universidad de Chile, Casilla 653, Santiago, Chile}

\begin{abstract}
The distribution of waiting times until the occurrence of a critical event is a crucial statistical problem across several disciplines in Science. 
In this work we present a statistical model in which a relevant quantity $X$ accumulates until overcoming a threshold $X^*$, which defines the collapse. The obtained waiting 
time distribution is a mixture of gamma distributions, which in turn can be approximated as an effective gamma distribution. 
\end{abstract}

\maketitle

\section{Introduction}
\label{introduction}

The study of the waiting time until the occurrence of some event is of crucial importance in reliability and engineering~\cite{Wolstenholme1999, Lawless2011}, nonequilibrium
statistical physics~\cite{Tang2014}, seismology~\cite{Davidsen2004} among other areas. In particular, gamma-distributed waiting times are relevant to several phenomena. They
have been observed, for instance, in earthquakes~\cite{Bak2002}, in modelling survival data~\cite{Gross1975}, stochastic processes in finance~\cite{Jagielski2017}, to name a few.

Several classes of phenomena have an interesting common theme: they can be understood in terms of the gradual accumulation of some quantity 
(e.g. stress in the case of earthquakes, cellular damage in the case of death of an organism~\cite{Belikov2017}, social unrest in the case of an uprising), where the collapse is triggered when 
some threshold is reached. If this increment is really gradual but unrelenting, the colloquial metaphor of the ``last straw that broke the camel's back'' may be quite appropriate~\cite{BBC2019}.

The phenomenon of collapse by gradual accumulation is also relevant to technological applications beside reliability, for instance, in the study of damage in structural 
materials~\cite{Caturla2000} and biological matter~\cite{Rube2010} induced by radiation.

In this work we present a general model of collapse due to incremental accumulation, which produces a waiting-time distribution in terms of a modified Bessel function, that however
can be approximated quite accurately by a gamma distribution. 

The rest of the paper is organized as follows. In Sections \ref{sect_model} and \ref{sect_approx}, we define our model and describe a simpler approximation. We finally close with some concluding remarks in Section \ref{sect_concluding}.

\section{A statistical model for collapse times}
\label{sect_model}

We will consider a statistical process where a positive quantity $X$ is accumulated from zero on incremental steps, until $X$ overcomes a threshold value $X^*$ which triggers the collapse. On each trial step (of fixed duration $\Delta t$), an attempt to increase the value of $X$ is performed with success probability $p$, that is, these attempts form a
Bernoulli process. On every successful trial, $X$ increases by an amount $\Delta X$ which may be fixed or random.

\begin{figure}[h!]
\begin{center}
\includegraphics[width=8cm]{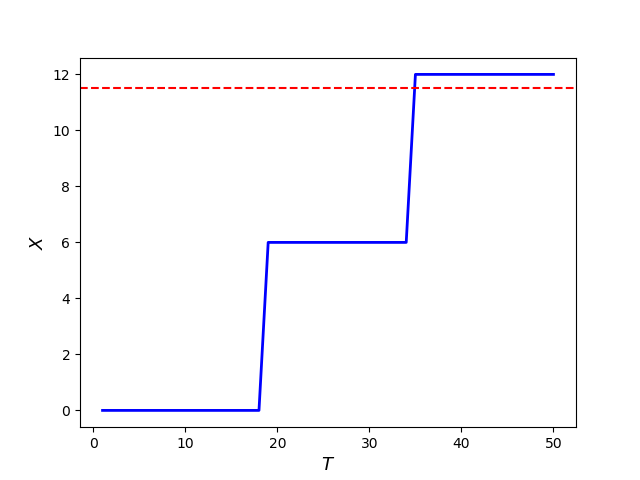}
\end{center}
\caption{Representation of the statistical process leading to the system collapse for the case where the increment $\Delta X$ is a positive constant. The dotted line corresponds to the value of the threshold $X^{*}$.}
\label{x_const}
\end{figure}

\begin{figure}[h!]
\begin{center}
\includegraphics[width=8cm]{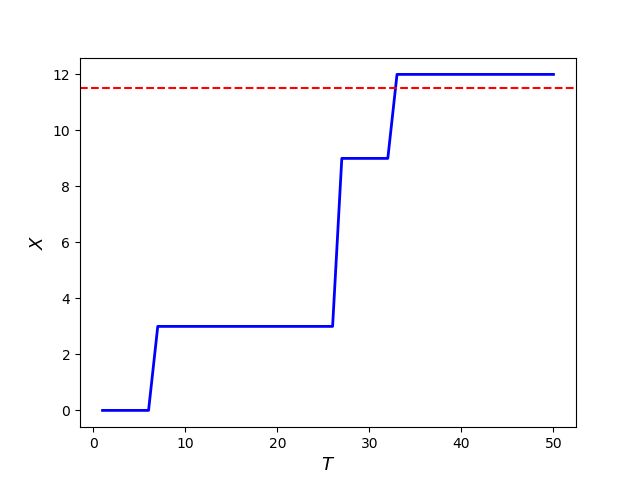}
\end{center}
\caption{Representation of the statistical process leading to the system collapse for the case where $X$ accumulates on each step $\Delta t$ by random, exponentially-distributed increments $\Delta X$. The dotted line corresponds to the value of the threshold $X^{*}$.}
\label{x_random}
\end{figure}

In the limit of $p\rightarrow 0$ and $\Delta t \rightarrow 0$ keeping the ratio $\Delta t/p$ finite, this Bernoulli process becomes a Poisson process~\cite{Ross1996}, in which the
elapsed time $\Delta T$ between successful attempts is exponentially distributed,
\begin{equation}
P(\Delta T|p, \Delta t) = \frac{1}{\tau}\exp(-\Delta T/\tau).
\label{P_t}
\end{equation}

\noindent
Here we have defined the \emph{characteristic time}
\begin{equation}
\tau := \frac{\Delta t}{p}
\end{equation}
for convenience, which corresponds to the main elapsed time between events, 
\begin{equation}
\tau = \big<\Delta T\big>_{p,\Delta t}.
\label{eq_tau}
\end{equation}

Now we will analyze the statistical process by considering two cases: (1) $\Delta X$ is a positive constant quantity, as shown in Fig. \ref{x_const}, and (2) $\Delta X$ is a 
positive quantity distributed exponentially, as shown in Fig. \ref{x_random}. In both cases the quantity $X$ accumulates from zero in incremental steps and can be analyzed by considering the number of increments $n_c$ needed for crossing the $X^*$ threshold that triggers the collapse. After $n$ accumulation events, the total elapsed time and the total amount of $X$ are given by
\begin{eqnarray}
T_n & := \sum_{i=1}^n {(\Delta T)}_i,
\label{eq_sumT} \\
X_n & := \sum_{i=1}^n {(\Delta X)}_i,
\label{eq_sumX}
\end{eqnarray}
respectively. This is in fact an instance of a continuous-time random walk (CTRW)~\cite{Montroll1965, Klafter2011}. We define the \emph{critical number of events} $n_c$ by the condition,
\begin{align}
X_{n_c-1} & < X^*, \nonumber \\
X_{n_c} & \geq X^*.
\label{eq_nc1}
\end{align}

Because of Eqs. \ref{P_t} and \ref{eq_sumT}, the quantity $T_{n}$ is the sum of $n$ i.i.d exponential variables which is gamma-distributed~\cite{Wolstenholme1999} with shape parameter $k=n$,
\begin{equation}
P(T_{n}=t_{w}|\tau) = \frac{\exp(-t_{w}/\tau) (t_{w})^{n-1}}{\tau^{n} \Gamma(n)}.
\label{eq_probTn}
\end{equation}

At this point the distribution of waiting times can be written formally as,
\begin{equation}
P(t_w|I, \tau) = \sum_{n=1}^{\infty} P(n_c = n|I) \cdot P(T_n = t_w|\tau),
\label{P_tw}
\end{equation}
which is a mixture of gamma distributions that can be interpreted as a superstatistical model~\cite{Beck2004} with uncertainty about the shape parameter of the gamma distribution 
in Eq. \ref{eq_probTn}. In the following sections we will give details on the procedure to compute $P(n_c|I)$ mentioned above.
\subsection{$\Delta X$ a positive constant}

We first consider the case where $\Delta X$ is a positive constant. After every interval $\Delta T$ the increment $\Delta X$ has a single possible 
value, so
\begin{equation}
P(\Delta X = \xi|I) = \delta(\Delta X - \xi),
\end{equation}
and therefore
\begin{equation}
X_{n} = \sum_{i=1}^n (\Delta X)_i = \sum_{i=1}^n \Delta X = n\;\Delta X,
\end{equation}
The sum of $n$ successive increments $X_{n}$ is clearly given by a Dirac delta distribution,
\begin{equation}
P(X_{n}=X|\Delta X) = \delta(X-n\;\Delta X).
\label{P_X}
\end{equation}

\noindent
From the definition of $n_c$ in Eq. \ref{eq_nc1} we can make use of Bayes's theorem as
\begin{equation}
P(n_c|\mathcal{C}, I) = \frac{P(n_c|I) \times P(\mathcal{C}|n_c, I)}{P(\mathcal{C}|I)},
\label{Bayes}
\end{equation}

\noindent
where $\mathcal{C}$ denotes the proposition
\begin{equation}
(r_{n_c-1} < 1) \wedge (r_{n_c} \geq 1),
\end{equation}

\noindent
with $r_n := X_n/X^*$ and
\begin{align}
P(\mathcal{C}|n_c, I) & = \int dr\;d\Delta X \:P(r_{n-1}=r|X^*, \Delta X) \cdot P(\Delta X|I) \nonumber \\
                      & \times \Theta(1-r)\Theta\left(\left[r+\frac{\Delta X}{X^*}\right]-1\right),
\label{eq_PCn}
\end{align}
where
\begin{equation}
P(r_n = r|X^*, \Delta X)  = \int_{0}^{\infty} dX\;P(X_n=X|\Delta X)\;\delta(r-r_n).
\end{equation}

The probability distribution $P(X_n|\Delta X)$ is given by Eq. \ref{P_X}, so $P(r_{n-1}=r|X^*, \Delta X)$ reduces to
\begin{align}
P(r_{n-1} = r|X^*, \Delta X) & = X^* \delta(rX^* - (n-1)\Delta X) \nonumber \\
                             & = \delta\left(r - (n-1)\left[\frac{\Delta X}{X^*}\right]\right).
\end{align}

\noindent
Replacing into Eq. \ref{eq_PCn}, we get
\begin{equation}
P(\mathcal{C}|n, I) = \Theta\Big(1 - (n-1)\left[\frac{\Delta X}{X^*}\right]\Big)\Theta\Big(n\left[\frac{\Delta X}{X^*}\right] - 1\Big).
\label{eq_PCn_2}
\end{equation}

\noindent
We see that these conditions admit a single value of $n$ that is between $$\frac{X^*}{\Delta X} \leq n < \frac{X^*}{\Delta X}+1,$$
therefore $P(\mathcal{C}|n, I) = 1$ if $n=\text{ceil}\left(\frac{X^*}{\Delta X}\right)$, and is zero otherwise. Then Eq. \ref{eq_PCn_2} is reduced to
the following expression
%
\begin{equation}
P(\mathcal{C}|n, I) = \delta(n, n_c)
\label{Dist_Pdelta}
\end{equation}
where $n_c=\text{ceil}(X^*/\Delta X)$. By Bayes's theorem, choosing a flat prior $P(n_c|I)=p_0$, we have also
\begin{equation}
P(n_c = n |X^*, \Delta X)=\delta(n, \text{ceil}(X^*/\Delta X))
\end{equation}
which, by replacing into Eq. \ref{P_tw} shows that the waiting times until the collapse of the system follow a gamma distribution for constant increments 
$\Delta X$,
\begin{equation}
P(t_w|n_c, \tau) = \frac{\exp(-t_w /\tau)(t_w)^{n_{c}-1}}{\tau^{n_c}(n_c-1)!},
\end{equation}
where we have replaced $\Gamma(n_c)=(n_c-1)!$. This gamma distribution with integer shape parameter is also known as the Erlang distribution~\cite{Ibe2014}. This is a well-known 
distribution in reliability studies, however it has also been found in studies on discrete carcinogenic events~\cite{Belikov2017}.

As is well known for the gamma distribution, the mean and variance are given by
\begin{equation}
\big<t_w\big>_{n_c, \tau} = \tau \cdot n_c,
\end{equation}
and
\begin{equation}
\big<(\delta t_w)^2\big>_{n_c, \tau} = \tau^2 \cdot n_c,
\end{equation}
respectively.

\subsection{$\Delta X$ from an exponential distribution}

In this case, we consider $\Delta X$ as an exponentially distributed variable,
\begin{equation}
P(\Delta X|\lambda) = \lambda \exp(-\lambda \Delta X),
\label{eq_prob_deltax}
\end{equation}
which can be justified from knowledge of the mean increment $\left<\Delta X\right>$ and the maximum entropy principle~\cite{Jaynes1957}.

We use the same Eq. \ref{eq_nc1} to describe the critical number of accumulation events $n_c$. However, now the quantity $X_n$ is, just as $T_n$, a sum of $n$ i.i.d exponential variables, hence it is also gamma-distributed with shape parameter $k=n$. We have then, from Eqs. \ref{eq_prob_deltax} and \ref{P_t}, that
\begin{equation}
P(X_n = X|\lambda) = \frac{\lambda^n\exp(-\lambda X_n)(X_n)^{n-1}}{\Gamma(n)}.
\label{eq_Xn}
\end{equation}

From Eq. \ref{eq_Xn} we can see that the ratio $r_n = X_n/X^*$ is also a gamma-distributed variable,
\begin{equation}
P(r_n = r|\alpha) = \frac{\exp(-r\alpha)r^{n-1}}{\Gamma(n)\alpha^{-n}},
\label{eq_Xn_ratio}
\end{equation}
with $\alpha$ the \emph{limit of collapse}, defined by
\begin{equation}
\alpha := \lambda X^* = \frac{X^*}{\big<\Delta X\big>_\lambda},
\end{equation}
the last equality because of the exponential distribution in Eq.~\ref{eq_prob_deltax}. From the definition of $n_c$ in Eq. \ref{eq_nc1} we can make use of Bayes's theorem in Eq. \ref{Bayes} as
\begin{align}
P(\mathcal{C}|n_c, I) & = \int\vspace{-7pt}dr \: d\Delta X_n\:P((r_{n-1}=r)\wedge r_n|\alpha)\times\nonumber\\
                      & \:\Theta(1-r)\Theta(r_n -1) \nonumber \\
                      & = \int\vspace{-7pt}dr\:d\Delta X_n\:P(r|\alpha)P(\Delta X_n|\lambda)\times\nonumber \\
                      & \Theta(1-r)\Theta\left(\left[r+\frac{\Delta X_n}{X^*}\right]-1\right) \nonumber \\
                      & = \int_0^1\vspace{-7pt}dr\:\frac{\exp(-r\alpha)r^{n-2}}{\Gamma(n-1)\alpha^{-n+1}}\times\nonumber\\
                      & \int_{X^{*}(1-r)}^{\infty}\vspace{-7pt}\hspace{-10pt}d\Delta X_n \:\lambda \exp(-\lambda\Delta X_n),
\end{align}
which reduces to
\begin{equation}
P(\mathcal{C}|n_c, I) = \frac{\exp(-\alpha)\alpha^{n-1}}{\Gamma(n)}.
\end{equation}

Now assuming a constant prior probability $P(n_c|I)=p_0$ we finally obtain
\begin{equation}
P(n_c = n|\alpha) = \frac{\exp(-\alpha)\alpha^{n-1}}{\Gamma(n)},
\label{eq_prob_nc}
\end{equation}
because the normalization constant $P(\mathcal{C}|I)$ is directly given by
\begin{equation}
P(\mathcal{C}|I) = \sum_{k=1}^\infty \frac{\exp(-\alpha)\alpha^{n-1}}{\Gamma(n)} = \sum_{k=1}^{\infty} \frac{\exp(-\alpha)\alpha^{n-1}}{(n-1)!} = 1.
\end{equation}

Eq. \ref{eq_prob_nc} is actually a Poisson distribution for the variable $n_c - 1$, hence 
\begin{equation}
\big<n_c\big>_\alpha = \alpha + 1.
\label{eq_nc_alpha}
\end{equation}
The comparison between $$\left<n_c\right> = \frac{X^*}{\left<\Delta X\right>} + 1$$
and the relation $n_c=\text{ceil}(X^*/\Delta X)$ for the earlier case seems quite illustrative. Eq. \ref{P_tw} then reduces to our main result,
\begin{align}
P(t_w|\alpha, \tau) &= \sum_{n=1}^\infty \frac{\exp(-\alpha)\alpha^{n-1}}{\Gamma(n)} \frac{\exp(-t_w/\tau)t_w^{n-1}}{\Gamma(n)\tau^n}\nonumber \\
                    &= \frac{\exp(-\alpha - t_w/\tau)}{\tau}I_0\left(2\sqrt{\frac{\alpha t_w}{\tau}}\right),
\label{eq_model_super}
\end{align}
where $I_0$ corresponds to the zero-order modified Bessel function of the first kind. This model is a particular case of a general purpose distribution reported originally by 
Laha~\cite{Laha1954, McNolty1967}, in contexts other than waiting times.

\begin{figure}[t!]
\begin{center}
\includegraphics[width=8cm]{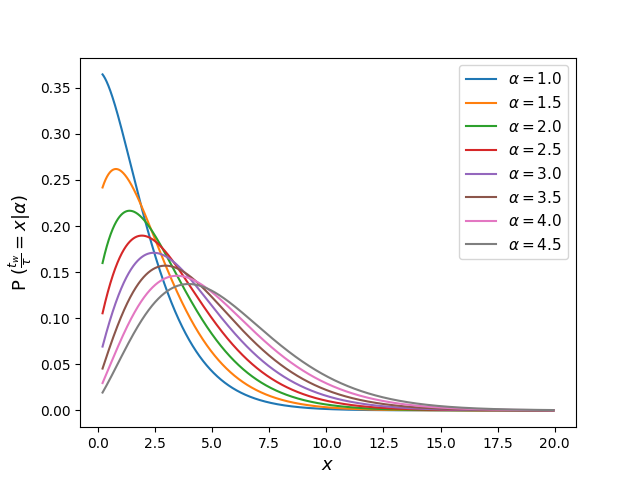}
\end{center}
\caption{Distribution of normalized waiting times $t_w/\tau$ (Eq. \ref{eq_model_super}) for different values of the shape parameter $\alpha$.}
\label{p_twtau}
\end{figure}
In this model, $\alpha$ takes the role of a shape parameter, while $\tau$ is clearly a scale parameter. Accordingly, we show in Fig. \ref{p_twtau} the behavior of the 
distribution of the normalized waiting time $t_w/\tau$ in Eq.\ref{eq_model_super} for different values of $\alpha$.

\noindent
The maximum likelihood equations for $P(t_w|\alpha, \tau)$ are, for a sample $t_1, \ldots, t_n$,
\begin{align}
\tau & = \frac{\overline{t}}{\alpha + 1}, \\
	\sqrt{\frac{\alpha}{\alpha + 1}} & = \left[\sqrt{t^{*}}\frac{I_{1}(2\sqrt{\alpha(\alpha +1)t^{*}})}{I_{0}(2\sqrt{\alpha(\alpha +1)t^{*}})}\right],
\end{align}
where $\overline{t}=\frac{1}{n}\sum_{i=1}^{n}t_{i}$, $t^{*}_{i}=\frac{t_{i}}{\overline{t}}$, $I_1(x)$ corresponds to the first-order modified Bessel function of the first kind, and 
$[f] := \frac{1}{n}\sum_{i=1}^{n}f_{i}$. The mean and variance of the distribution are given by
\begin{equation}
\big<t_w\big>_{\alpha, \tau} = \tau(\alpha + 1)
\end{equation}
and
\begin{equation}
\big<(\delta t_w)^2\big>_{\alpha, \tau} = \tau^{2}(1+2\alpha),
\end{equation}
respectively. Note that, by using Eq. \ref{eq_tau} and \ref{eq_nc_alpha}, the mean value of $t_w$ can also be expressed as
\begin{equation}
\big<t_w\big>_{\alpha, \tau} = \left<\Delta T\right>_\tau \cdot \left<n_c\right>_\alpha.
\end{equation}

\nocite{Conolly1979} 
\section{An empirical approximation}
\label{sect_approx}

\begin{figure}[h!]
\begin{center}
\includegraphics[width=8cm]{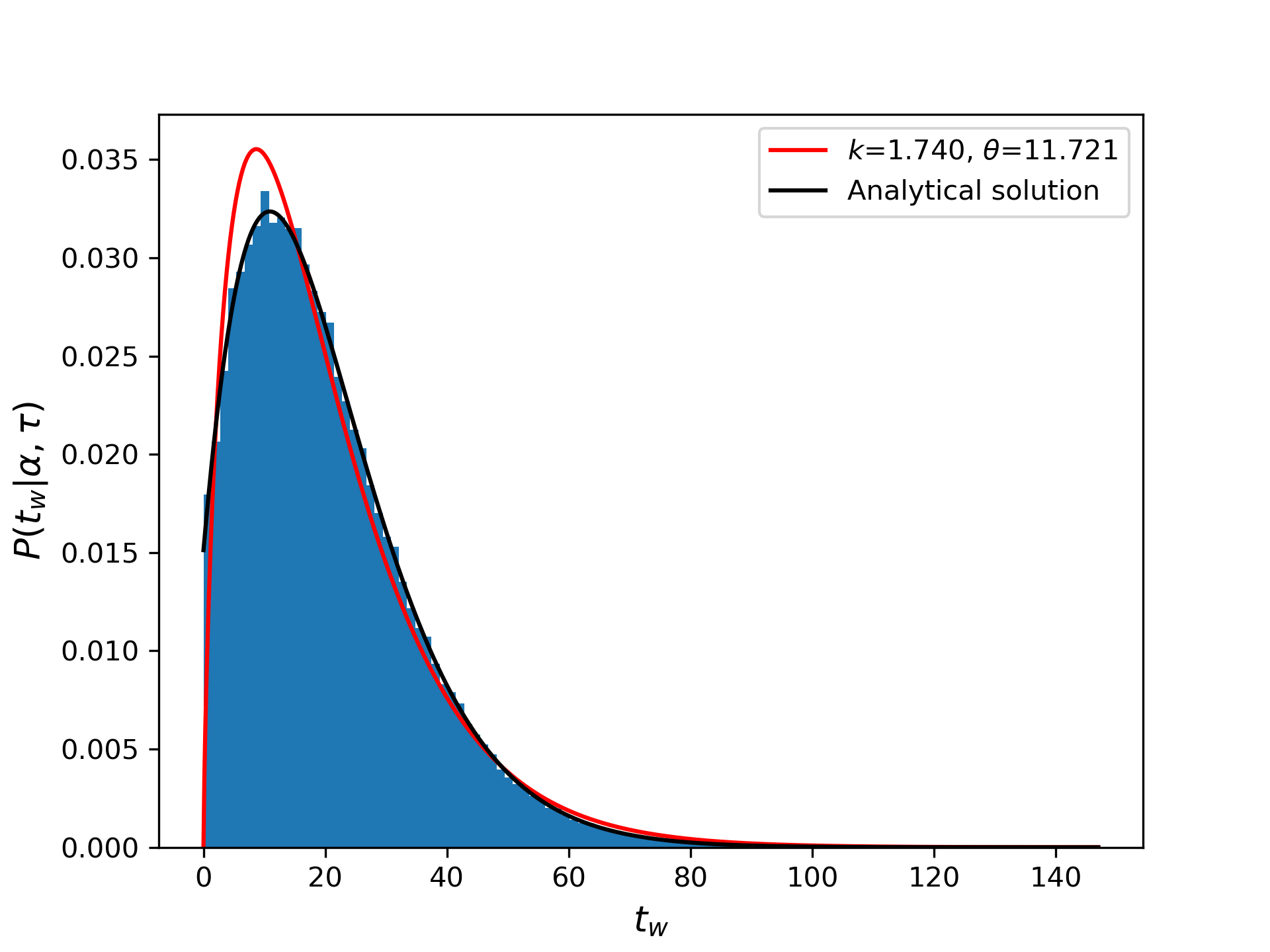}
\end{center}
\caption{Histogram of waiting times obtained by direct numerical sampling, together with the analytical model of Eq. \ref{eq_model_super} for $\alpha$=2.4, $\tau$=6 (black line) and a gamma model (red line).}
\label{fig_histo_simple}
\end{figure}

Even when our main result, Eq. \ref{eq_model_super} is exact, it is intractable in statistical terms because the maximum likelihood equations cannot be written in terms of sufficient
statistics which are independent of the parameters. Despite this drawback, we have found empirically by direct numerical sampling of
Eqs. \ref{eq_Xn} and \ref{eq_probTn} that Eq. \ref{eq_model_super} can be very closely approximated by a gamma distribution, as shown in Fig. \ref{fig_histo_simple}.

\begin{figure}[h!]
\begin{center}
\includegraphics[width=8cm]{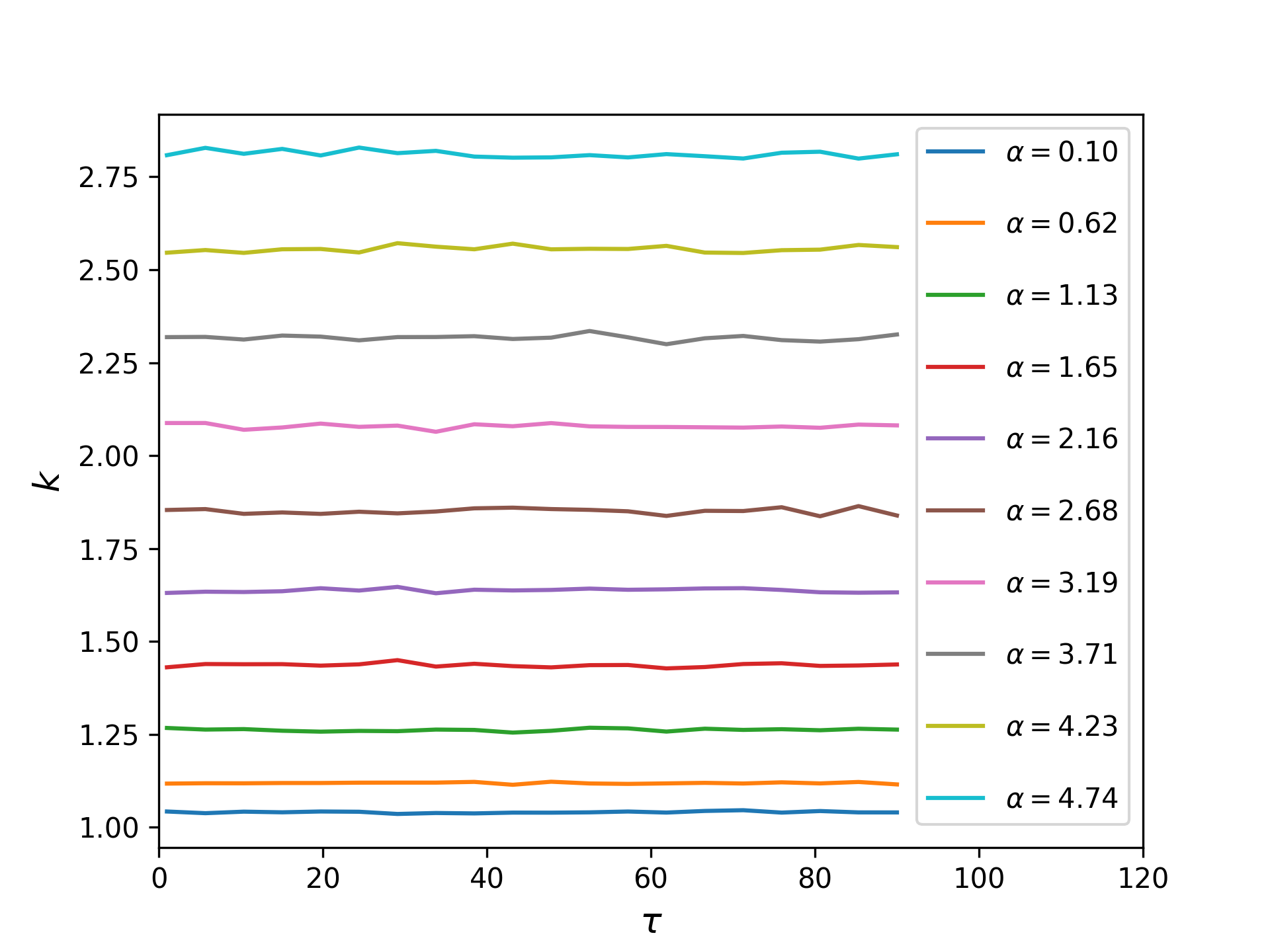}
\includegraphics[width=8cm]{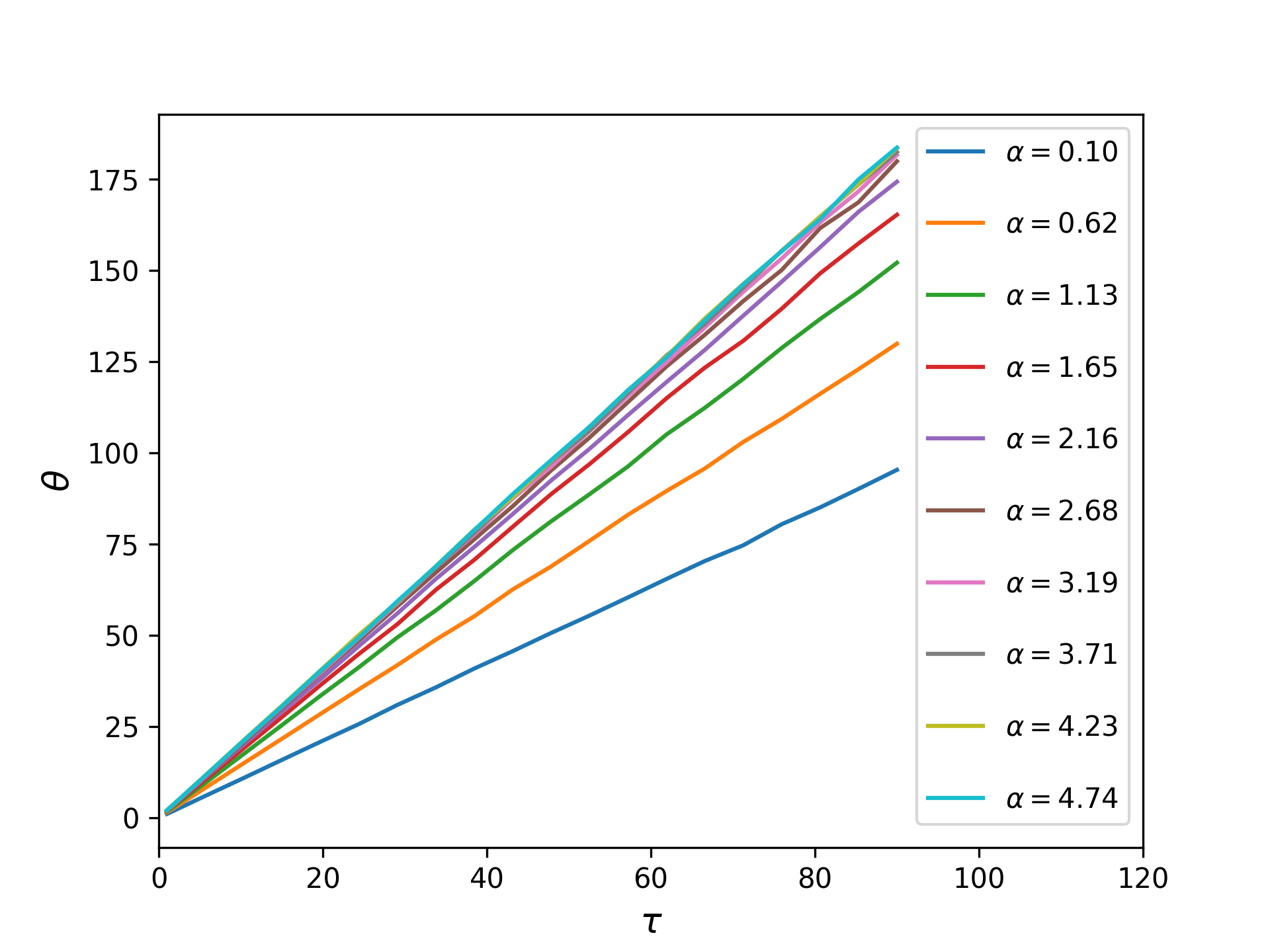}
\end{center}
\caption{Upper panel, shape parameter $k$ as a function of $\tau$ for different values of $\alpha$. Lower panel, scale parameter $\theta$ as a function of $\tau$
for different values of $\alpha$.}
\label{fig_k_theta}
\end{figure}

\begin{figure}[h!]
\begin{center}
\includegraphics[width=8cm]{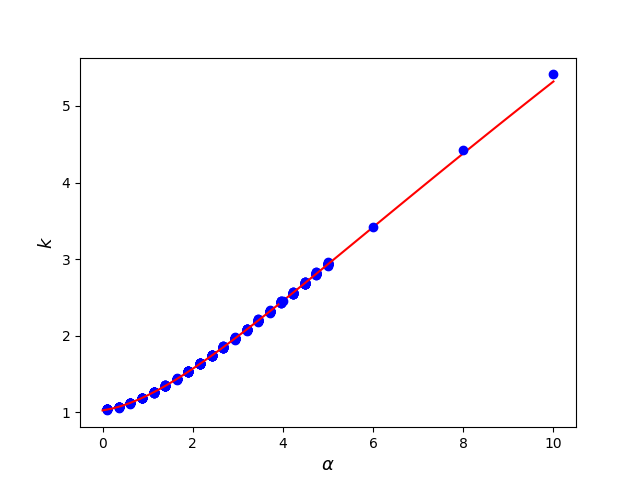}
\includegraphics[width=8cm]{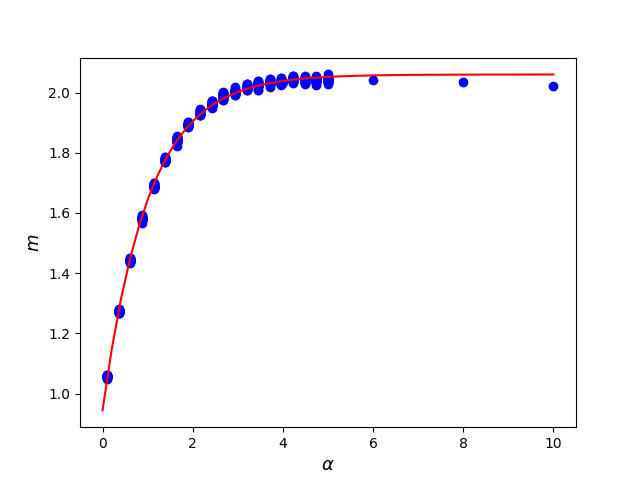}
\end{center}
\caption{Upper panel, $k(\alpha)$ as a function of $\alpha$. Lower panel, $m(\alpha)$ as a function of $\alpha$.}
\label{fig_k_m}
\end{figure}

From Fig. \ref{fig_k_theta} it is clear that this approximate gamma distribution has parameters
\begin{align}
k(\alpha, \tau) & = k(\alpha), \nonumber \\
\theta(\alpha, \tau) & = \tau\cdot m(\alpha),
\end{align}
where the universal functions $k(\alpha)$ and $m(\alpha)$ are given by
\begin{equation}
k(\alpha)=(A\alpha + B)\left[1-\exp(-C\alpha + D\alpha^{2})\right] + E,
\end{equation}
with $A=0.4582$, $B=0.2546$, $C=0.0116$, $D=0.0116$, $E=1.0249$, and
\begin{equation}
m(\alpha)=m_{0}\left[1-\exp(-\alpha/A)\right] + B,
\end{equation}
with $m_{0}=1.1161$, $A=1.0170$ and $B=0.944$, as shown in Fig. \ref{fig_k_m}.

It is important to note that, in the limit of $\alpha \rightarrow 0$ we have from Eq. \ref{eq_prob_nc} that $$\lim_{\alpha \rightarrow 0} P(n_c|\alpha) = \delta(n_c, 1),$$
hence the distribution of waiting times reduces to the exponential distribution in Eq. \ref{P_t}. This is clearly expected, as $\alpha \rightarrow 0$ corresponds to vanishing
threshold $X^* \rightarrow 0$ which induces an immediate crossing on the first attempt.

\section{Concluding remarks}
\label{sect_concluding}

We have developed a statistical model for the waiting time until collapse of a system, induced by gradual accumulation of some quantity, in a general context. 
We consider for illustration two main cases of interest, namely when the increment $\Delta X$ is a positive constant, in which case the waiting time distribution is an Erlang distribution, and another where $\Delta X$ is an exponential random variable, changing at each time step $\Delta t$. The latter case produces a waiting
time distribution in terms of the modified Bessel function $I_0$.

We have fitted an effective gamma distribution to our model $P(t_w|\alpha, \tau)$, from which we can see that the shape parameter $k$ does not depend on the 
characteristic time $\tau$ for fixed value of the limit of collapse $\alpha$. As $\tau$ is a scale parameter, and because $\big<t_w\big> = \tau(\alpha+1)$, the probability distribution 
of $t_w$ normalized by its mean only depends on the shape parameter $\alpha$, and therefore can describe universal behavior for a fixed $\alpha$. We have that, for different values of $\alpha$ the behavior of the Bessel function distribution in Eq. \ref{eq_model_super} (as well as the effective gamma distribution) changes, as seen in Fig. \ref{p_twtau}.

For small values of $\alpha$ the distribution is narrow, resembling an exponential distribution, and the waiting time before the collapse occurs is small, 
concentrated around zero. The number of steps required to cross the threshold decreases in $T$ as seen in Fig. \ref{x_random}. When 
$\alpha$ increases, the distribution resembles a Gaussian distribution, and the waiting time for the collapse to occur is longer, requiring therefore a greater number of steps in $T$ to reach the threshold, which means that when the quantity $\Delta X \ll X^{*}$, i.e. the value of $\alpha \gg 1$, the quantity $X$
accumulates enough after $n$ events occur for a time $T$ until the collapse occurs. However, when $\Delta X \approx X^{*}$, we have $\alpha \approx$ $1$, the number of events that occur is smaller, and therefore the waiting time to cross the threshold $X^{*}$ decreases.

\section*{Acknowledgements}

This work is supported by FONDECYT 1171127 and Anillo ACT-172101 grants.


\begin{thebibliography}{20}
\expandafter\ifx\csname natexlab\endcsname\relax\def\natexlab#1{#1}\fi
\expandafter\ifx\csname bibnamefont\endcsname\relax
  \def\bibnamefont#1{#1}\fi
\expandafter\ifx\csname bibfnamefont\endcsname\relax
  \def\bibfnamefont#1{#1}\fi
\expandafter\ifx\csname citenamefont\endcsname\relax
  \def\citenamefont#1{#1}\fi
\expandafter\ifx\csname url\endcsname\relax
  \def\url#1{\texttt{#1}}\fi
\expandafter\ifx\csname urlprefix\endcsname\relax\def\urlprefix{URL }\fi
\providecommand{\bibinfo}[2]{#2}
\providecommand{\eprint}[2][]{\url{#2}}

\bibitem[{\citenamefont{Wolstenholme}(1999)}]{Wolstenholme1999}
\bibinfo{author}{\bibfnamefont{L.~C.} \bibnamefont{Wolstenholme}},
  \emph{\bibinfo{title}{Reliability Modelling: A statistical approach}}
  (\bibinfo{publisher}{CRC Press}, \bibinfo{year}{1999}).

\bibitem[{\citenamefont{Lawless}(2011)}]{Lawless2011}
\bibinfo{author}{\bibfnamefont{J.}~\bibnamefont{Lawless}},
  \emph{\bibinfo{title}{Statistical Models and Methods for Lifetime Data}},
  Wiley Series in Probability and Statistics (\bibinfo{publisher}{Wiley},
  \bibinfo{year}{2011}).

\bibitem[{\citenamefont{Tang et~al.}(2014)\citenamefont{Tang, Xu, and
  Wang}}]{Tang2014}
\bibinfo{author}{\bibfnamefont{G.-M.} \bibnamefont{Tang}},
  \bibinfo{author}{\bibfnamefont{F.}~\bibnamefont{Xu}}, \bibnamefont{and}
  \bibinfo{author}{\bibfnamefont{J.}~\bibnamefont{Wang}},
  \bibinfo{journal}{Physical Review B} \textbf{\bibinfo{volume}{89}},
  \bibinfo{pages}{205310} (\bibinfo{year}{2014}).

\bibitem[{\citenamefont{Davidsen and Goltz}(2004)}]{Davidsen2004}
\bibinfo{author}{\bibfnamefont{J.}~\bibnamefont{Davidsen}} \bibnamefont{and}
  \bibinfo{author}{\bibfnamefont{C.}~\bibnamefont{Goltz}},
  \bibinfo{journal}{Geophysical Research Letters} \textbf{\bibinfo{volume}{31}}
  (\bibinfo{year}{2004}).

\bibitem[{\citenamefont{Bak et~al.}(2002)\citenamefont{Bak, Christensen, Danon,
  and Scanlon}}]{Bak2002}
\bibinfo{author}{\bibfnamefont{P.}~\bibnamefont{Bak}},
  \bibinfo{author}{\bibfnamefont{K.}~\bibnamefont{Christensen}},
  \bibinfo{author}{\bibfnamefont{L.}~\bibnamefont{Danon}}, \bibnamefont{and}
  \bibinfo{author}{\bibfnamefont{T.}~\bibnamefont{Scanlon}},
  \bibinfo{journal}{Physical Review Letters} \textbf{\bibinfo{volume}{88}},
  \bibinfo{pages}{178501} (\bibinfo{year}{2002}).

\bibitem[{\citenamefont{Gross and Clark}(1975)}]{Gross1975}
\bibinfo{author}{\bibfnamefont{A.~J.} \bibnamefont{Gross}} \bibnamefont{and}
  \bibinfo{author}{\bibfnamefont{V.~A.} \bibnamefont{Clark}},
  \emph{\bibinfo{title}{Survival distributions: reliability applications in the
  biomedical sciences}} (\bibinfo{publisher}{John Wiley and Sons},
  \bibinfo{year}{1975}).

\bibitem[{\citenamefont{Jagielski et~al.}(2017)\citenamefont{Jagielski, Kutner,
  and Sornette}}]{Jagielski2017}
\bibinfo{author}{\bibfnamefont{M.}~\bibnamefont{Jagielski}},
  \bibinfo{author}{\bibfnamefont{R.}~\bibnamefont{Kutner}}, \bibnamefont{and}
  \bibinfo{author}{\bibfnamefont{D.}~\bibnamefont{Sornette}},
  \bibinfo{journal}{Physica A} \textbf{\bibinfo{volume}{483}},
  \bibinfo{pages}{68} (\bibinfo{year}{2017}).

\bibitem[{\citenamefont{Belikov}(2017)}]{Belikov2017}
\bibinfo{author}{\bibfnamefont{A.~V.} \bibnamefont{Belikov}},
  \bibinfo{journal}{Scientific Reports} \textbf{\bibinfo{volume}{7}},
  \bibinfo{pages}{12170} (\bibinfo{year}{2017}).

\bibitem[{BBC(2019)}]{BBC2019}
\emph{\bibinfo{title}{Do today's global protests have anything in common?}},
  \bibinfo{howpublished}{\url{https://www.bbc.com/news/world-50123743}}
  (\bibinfo{year}{2019}).

\bibitem[{\citenamefont{Caturla et~al.}(2000)\citenamefont{Caturla, Soneda,
  Alonso, Wirth, Rubia, and Perlado}}]{Caturla2000}
\bibinfo{author}{\bibfnamefont{M.~J.} \bibnamefont{Caturla}},
  \bibinfo{author}{\bibfnamefont{N.}~\bibnamefont{Soneda}},
  \bibinfo{author}{\bibfnamefont{E.}~\bibnamefont{Alonso}},
  \bibinfo{author}{\bibfnamefont{B.~D.} \bibnamefont{Wirth}},
  \bibinfo{author}{\bibfnamefont{T.~D. D.~L.} \bibnamefont{Rubia}},
  \bibnamefont{and} \bibinfo{author}{\bibfnamefont{J.~M.}
  \bibnamefont{Perlado}}, \bibinfo{journal}{Journal of Nuclear Materials}
  \textbf{\bibinfo{volume}{276}}, \bibinfo{pages}{13} (\bibinfo{year}{2000}).

\bibitem[{\citenamefont{R{\"u}be et~al.}(2010)\citenamefont{R{\"u}be, Fricke,
  Wendorf, St{\"u}tzel, K{\"u}hne, Ong, Lipp, and R{\"u}be}}]{Rube2010}
\bibinfo{author}{\bibfnamefont{C.~E.} \bibnamefont{R{\"u}be}},
  \bibinfo{author}{\bibfnamefont{A.}~\bibnamefont{Fricke}},
  \bibinfo{author}{\bibfnamefont{J.}~\bibnamefont{Wendorf}},
  \bibinfo{author}{\bibfnamefont{A.}~\bibnamefont{St{\"u}tzel}},
  \bibinfo{author}{\bibfnamefont{M.}~\bibnamefont{K{\"u}hne}},
  \bibinfo{author}{\bibfnamefont{M.~F.} \bibnamefont{Ong}},
  \bibinfo{author}{\bibfnamefont{P.}~\bibnamefont{Lipp}}, \bibnamefont{and}
  \bibinfo{author}{\bibfnamefont{C.}~\bibnamefont{R{\"u}be}},
  \bibinfo{journal}{International Journal of Radiation Oncology* Biology*
  Physics} \textbf{\bibinfo{volume}{76}}, \bibinfo{pages}{1206}
  (\bibinfo{year}{2010}).

\bibitem[{\citenamefont{Ross}(1996)}]{Ross1996}
\bibinfo{author}{\bibfnamefont{S.~M.} \bibnamefont{Ross}},
  \emph{\bibinfo{title}{Stochastic processes}} (\bibinfo{publisher}{Wiley},
  \bibinfo{year}{1996}).

\bibitem[{\citenamefont{Montroll and Weiss}(1965)}]{Montroll1965}
\bibinfo{author}{\bibfnamefont{E.~W.} \bibnamefont{Montroll}} \bibnamefont{and}
  \bibinfo{author}{\bibfnamefont{G.}~\bibnamefont{Weiss}}, \bibinfo{journal}{J.
  Math. Phys.} \textbf{\bibinfo{volume}{6}}, \bibinfo{pages}{167}
  (\bibinfo{year}{1965}).

\bibitem[{\citenamefont{Klafter and Sokolov}(2011)}]{Klafter2011}
\bibinfo{author}{\bibfnamefont{J.}~\bibnamefont{Klafter}} \bibnamefont{and}
  \bibinfo{author}{\bibfnamefont{I.~M.} \bibnamefont{Sokolov}},
  \emph{\bibinfo{title}{First steps in random walks: from tools to
  applications}} (\bibinfo{publisher}{Oxford University Press},
  \bibinfo{year}{2011}).

\bibitem[{\citenamefont{Beck}(2004)}]{Beck2004}
\bibinfo{author}{\bibfnamefont{C.}~\bibnamefont{Beck}}, \bibinfo{journal}{Cont.
  Mech. Thermodyn.} \textbf{\bibinfo{volume}{16}}, \bibinfo{pages}{293}
  (\bibinfo{year}{2004}).

\bibitem[{\citenamefont{Ibe}(2014)}]{Ibe2014}
\bibinfo{author}{\bibfnamefont{O.}~\bibnamefont{Ibe}},
  \emph{\bibinfo{title}{Fundamentals of applied probability and random
  processes}} (\bibinfo{publisher}{Academic Press}, \bibinfo{year}{2014}).

\bibitem[{\citenamefont{Jaynes}(1957)}]{Jaynes1957}
\bibinfo{author}{\bibfnamefont{E.~T.} \bibnamefont{Jaynes}},
  \bibinfo{journal}{Physical Review} \textbf{\bibinfo{volume}{106}},
  \bibinfo{pages}{620} (\bibinfo{year}{1957}).

\bibitem[{\citenamefont{Laha}(1954)}]{Laha1954}
\bibinfo{author}{\bibfnamefont{R.~G.} \bibnamefont{Laha}},
  \bibinfo{journal}{Bulletin of the Calcutta Mathematical Society}
  \textbf{\bibinfo{volume}{46}}, \bibinfo{pages}{59} (\bibinfo{year}{1954}).

\bibitem[{\citenamefont{McNolty}(1967)}]{McNolty1967}
\bibinfo{author}{\bibfnamefont{F.}~\bibnamefont{McNolty}},
  \bibinfo{journal}{The Indian Journal of Statistics, Series B (1960-2002)}
  \textbf{\bibinfo{volume}{29}}, \bibinfo{pages}{235} (\bibinfo{year}{1967}).

\bibitem[{\citenamefont{Conolly and Choo}(1979)}]{Conolly1979}
\bibinfo{author}{\bibfnamefont{B.~W.} \bibnamefont{Conolly}} \bibnamefont{and}
  \bibinfo{author}{\bibfnamefont{Q.~H.} \bibnamefont{Choo}},
  \bibinfo{journal}{SIAM J. Appl. Math.} \textbf{\bibinfo{volume}{37}},
  \bibinfo{pages}{263} (\bibinfo{year}{1979}).

\end{thebibliography}

\end{document}